\documentclass[11pt]{elsart}
\usepackage{setspace}
\usepackage{bm}
\usepackage{cases}
\usepackage[dvips]{graphicx}

\begin{document}
\begin{frontmatter}

\title{Effects of awareness diffusion and self-initiated awareness behavior on epidemic spreading - an approach based on multiplex networks}
\author{Jia-Qian Kan$^{a}$}
\author{Hai-Feng Zhang$^{a,b,c}$}\footnote{Corresponding author:
haifengzhang1978@gmail.com}

\address
{$^{a}$ School of Mathematical Science, Anhui University, Hefei
230601, P. R. China }

\address{$^{b}$  Research centre of information supply \& assurance, Anhui University, Hefei
230601, P. R. China }
\address {$^{c}$ Department of Communication Engineering, North University of China, Taiyuan, Shan'xi 030051, P. R. China}

\begin{abstract}


In this paper, we study the interplay between the epidemic spreading
and the diffusion of awareness in multiplex networks. In the model,
an infectious disease can spread in one network representing the
paths of epidemic spreading (contact network), leading to the diffusion of awareness in the other network (information network), and then the diffusion
of awareness will cause individuals to take social distances, which in turn affects the epidemic spreading. As for the diffusion of awareness, we
assume that, on the one hand, individuals can be informed by other aware neighbors in information network, on the other hand, the susceptible individuals can be
\emph{self-awareness} induced by the infected neighbors in the
contact networks (local information) or mass media (global
information). Through Markov chain approach and numerical
computations, we find that the density of infected individuals and the
epidemic threshold can be affected by the structures of the two
networks and the effective transmission rate of the
awareness. However, we prove that though the
introduction of the \emph{self-awareness} can lower the
density of infection, which cannot increase the epidemic threshold
no matter of the local information or global information. Our
finding is remarkably different to many previous results--local information
based behavioral response can alter the epidemic threshold.

\end{abstract}

\date{}
\end{frontmatter}

\section{Introduction} \label{sec:intro}

The outbreaks of diseases can involve the diffusion of information in regard to the diseases, including the risk of
infection, rumors, fears and so on, which can stimulate
individuals to take spontaneous behavioral responses to protect
themselves, thereby bring profound impacts on the spreading of
disease~\cite{funk2009spread,kitchovitch2010risk,zhang2014effects,bu2013efficient,funk2010modelling,yang2012efficient}. For example, recent outbreaks of the H1N1 flu, the bird flu, and the severe acute respiratory syndrome
(SARS) have brought the reduction of going out and the plenty of
people wearing face masks. For
this reason, there has been an increasing focus on the development
of formal models aimed at investigating the interplay of epidemic
spreading and information-based behavioral responses\cite{bagnoli2007risk,wang2013impact,wang2013human,ruan2012epidemic}. Such as, based
on the assumption that the probability of susceptible individual
going to the alter state is proportional to the number of infected
neighbors, Sahneh\emph{et al.} extended the SIS
(Susceptible-Infected-Susceptible) model to a
Susceptible-Alter-Infected-Susceptible (SAIS) model~\cite{sahneh2011epidemic,sahneh2012existence},
and they found that the way of behavioral response can enhance the
epidemic threshold; Meloni \emph{et~al.} studied a meta-population
model that incorporates several scenarios of self-initiated
behavioral changes into the mobility patterns of individuals, and
they found that such behavioral changes do not alter the epidemic
threshold, but may produce a negative impact on disease, i.e., the
density of infection is increased~\cite{meloni2011modeling}; In Refs.~\cite{wu2012impact,wu2014responsive,zhang2014suppression}, authors
investigated the effects of the information-based behavioral
responses on the epidemic dynamics by designing the transmission
rate as a function of the local infected density or the global
infected density.

Though the effects of information-based behavioral responses on the
epidemic dynamics have been studied by many authors, most of works
assumed the spreadings of information and epidemic are in the same
network. As we know, with the development of technology, information
can fast diffuse through many different channels, such as, the word
of mouth, news media, online social networks, and so on. In view of
this, recent well-studied multiplex network theory has been used
to mimic the interplay of information or related awareness and the
epidemic dynamics\cite{gomez2013diffusion,wang2013interdependent,de2013mathematical,boccaletti2014structure}.
For instance, Sahneh \emph{et al.} have shown
that the information dissemination spread in another network can
help boost the resilience of the agents' population against the
spreading and found optimal information dissemination for different
topologies~\cite{sahneh2012optimal}; Wang \emph{et al.} investigated the
interplay of the epidemic dynamics and the information dynamics in
multiplex network based on the SIR (R-recovery) model, and focused on
the two fundamental quantities underlying any spreading process:
epidemic threshold and the final epidemic prevalence~\cite{wang2014asymmetrically}; Granell
\emph{et~al.} established an SIS-UAU model to investigate the
competing effects of the spreading of awareness and the epidemic
dynamics in multiplex with the transmission rate of awareness as
well as the structure of information network~\cite{granell2013dynamical}. More recently, they
further generalized their model by reducing the probability of
infected individuals becoming awareness and including the effect
of a mass broadcast of awareness (mass media) on the epidemic
dynamics~\cite{granell2014competing}.

Existing works either assume that individuals are self-initiated~\cite{meloni2011modeling,sahneh2012existence},
that is, they become aware because their neighbors are infected, or
individuals can only be informed and become aware by other aware
neighbors~\cite{wang2014asymmetrically,granell2013dynamical}, the combine effects of the two factors have not been well
studied. In reality, individuals can obtain the disease information through many ways. Inspired these factors, in the current work, we study the
interplay between the diffusion of awareness by incorporating the
\emph{self-awareness} effects and the epidemic dynamics under the
framework of multiplex network. In the model, an infectious disease
first spreads among population represented by the contact network,
and then the outbreak of the disease stimulates some people
(infected or susceptible individuals) become aware of the risk of
infection, and they take some protections to reduce the probability of
infection. Meanwhile, unaware individuals can be informed by other
aware individuals through the information network or become \emph{self-awareness}
induced by the infected neighbors in contact network or mass media. The finding
indicates that the additional self-initiated awareness mechanism can
reduce the density of infection, however, which can not alter the epidemic threshold. The results are
verified by the Monte-Carlo simulations and the microscopic Markov
chain approach (MMCA).

The layout of the paper is as follows: we introduce the model in
Sec.~\ref{sec:model}. The simulation results and theoretical
analysis are presented in Sec.~\ref{sec:main results}. Finally,
Conclusions and discussions are presented in
Sec.~\ref{sec:discussion}. The results for the global information-based \emph{self-initiated } awareness are given in Appendix.

\section{Model} \label{sec:model}

In this work, we generalize the model of Ref.~\cite{granell2013dynamical}. In that model, a
multiplex includes two layers, one is physical layer representing the spreading of epidemic (contact network), and the other is
information layer on where the diffusion of the awareness evolves
(information network). All nodes represent the same individuals in
both layers, but the connectivity is different in each of them. In
the contact layer, a Susceptible - Infected - Susceptible (SIS)
model is used to mimic the epidemic dynamics. That is to say, a
susceptible node can be infected by one infected neighbor with
certain probability, and the infected node can return to susceptible
state with probability $\mu$. On the information layer, the
dynamical process of awareness is assumed to be similar to the SIS
model, that is, an unaware node (U) can be informed by an aware
neighbor (A) with probability $\lambda$, and the aware node can loss
awareness and back to unaware state with probability $\delta$. The
interplay of the two processes is modelled as follows: once an
individual is infected, s/he will certainly become aware, that is,
the probability is $\sigma=100\%$.  In addition, to distinguish the
protective behaviors between the aware individuals and unaware individuals, let
$\beta$ and $\beta^{A}=\gamma\beta$ (here $0\leq\gamma<1$. If
$\gamma=0$, the aware individuals are completely immune to the
infection.) be the probabilities of unaware and aware susceptible
nodes to get infected, respectively.

From the description of the model, one can find that, on the one hand,
the authors assumed that the infected individuals will
\emph{automatically} become aware and are willing to inform the disease information. As we know, in many cases, infected individuals are
unwilling to tell others since they can be discriminated or isolated
by others once others know they are infected by one certain disease. So we make a progress and
assume that infected individuals becoming aware with probability
$0\leq\sigma\leq1$. On the other hand, in the model, individuals can
\emph{only} be informed by their neighbors through the information
network, that is to say, one individual has no chance to become aware
once their neighbors are unaware. However, as proposed in many
previous works, individuals can become \emph{self-awareness} once their friends are
infected or they are informed by the mass media. Thus, we further
generalize the model as: a susceptible individual can go to aware
state by \emph{self-initiated} response with probability $\kappa$
when contacting one infected friend. Therefore, the probability of the susceptible
individual becoming awareness increases with the number of infected
neighbors in the contact network~\cite{sahneh2012existence}. Note that, for the
original model in Ref.~\cite{granell2013dynamical},  awareness cannot break out if the probability $\sigma=0$, so the roles of awareness cannot be played,
however, in our model, the awareness can diffuse among population
even though $\sigma=0$ since susceptible individuals can become awareness by their
\emph{self-initiated} responses. When $\kappa=0.0$
and $\sigma=1.0$, our model returns to the original model in
Ref.~\cite{granell2013dynamical}.

\begin{figure}
\begin{center}
\includegraphics[height=100mm,width=100mm]{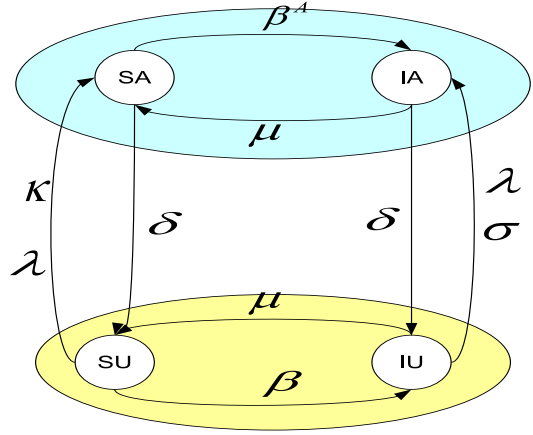}
\caption{Model description for the UAU-SIS dynamic. An individual can be in four
different states: SU, SA, IU, and IA. The top (bottom) layer is the epidemic process for the aware (unaware) individuals, respectively.
SU (SA) can be infected by an infectious neighbor in
contact layer with a probability $\beta$ ($\beta^{A}=\gamma\beta$).
IU and IA recovers to SU and SA, respectively, with the same
probability $\mu$; The left (right) flow is the awareness process for the susceptible (infected) individuals, respectively.
SU can go to SA  with a probability
$\lambda$ of being informed by an aware neighbor through information network, or induced by the infected neighbors in contact network
with a probability $\kappa$. SA recovers to SU with a probability
$\delta$. IU can go to IA by informed the aware neighbors
in information layer with a probability $\lambda$, or self-awareness
with a probability $\sigma$. SA and IA can become unaware and return
to SU and IU with  the same probability $\delta$. }\label{fig1}
\end{center}
\end{figure}

 According to this scheme, an individual can
be in four different states: susceptible and unaware(SU),
susceptible and aware(SA),infected and unaware(IU), infected and
aware(IA). The flow diagram of the model is given in
Fig.~\ref{fig1}.

\section{Main results} \label{sec:main results}
\subsection{Theoretical analysis}\label{analysis}

Denoting $a_{ij}$ and $b_{ij}$ be the adjacency matrices that support the SIS and UAU processes, respectively.
The probability of $i$ in one of four states at time $t$ is denoted
by $p_{i}^{SU}(t)$, $p_{i}^{SA}(t)$, $p_{i}^{IU}(t)$ and
$p_{i}^{IA}(t)$ respectively. Assuming the probability of susceptible
(infect) node $i$ \emph{not} being \emph{informed} by any neighbors is
$\theta_{i}(t)$ ($r_{i}(t)$), and the probability of unaware
(aware) susceptible node $i$ \emph{not} being \emph{infected} is
$q_{i}^{U}(t)$ ($q_{i}^{A}(t)$). They are described as~\cite{granell2013dynamical}:
\begin{eqnarray}
    \theta_{i}(t)&=&\prod_{j}(1-b_{ji}p_{j}^{A}(t)\lambda)(1-a_{ji}p_{j}^{I}(t)\kappa),\label{1}\\
    r_{i}(t)&=&\prod_{j}(1-b_{ji}p_{j}^{A}(t)\lambda),\label{2}\\
    q_{i}^{U}(t)&=&\prod_{j}(1-a_{ji}p_{j}^{I}(t)\beta),\label{3}\\
    q_{i}^{A}(t)&=&\prod_{j}(1-a_{ji}p_{j}^{I}(t)\beta^{A}),\label{4}
\end{eqnarray}
where $p_{j}^{A}(t)=p_{j}^{SA}(t)+p_{j}^{IA}(t)$ and
$p_{j}^{I}(t)=p_{j}^{IU}(t)+p_{j}^{IA}(t)$ (Note: to simplify the
model, we do not distinguish the infectivity of IA and IU,
meanwhile, the diffusion capabilities of awareness for SA and IA are
also the same.).
\begin{figure}
\begin{center}
\includegraphics[height=120mm,width=120mm]{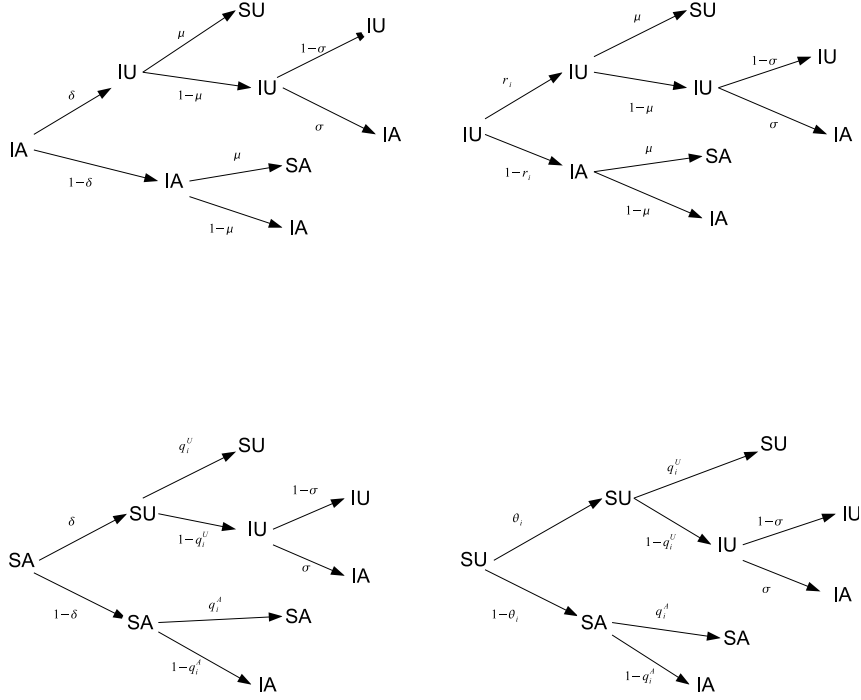}
\caption{Transition probability trees for the states of UAU-SIS
dynamics in the multiplex for per time step. The denotations of $\theta_i$, $r_i$, $q^{U}_i$ and  $q^{A}_i$ are given in Eqs.~(\ref{1})-(\ref{4}). Other parameters have the same denotations as in Fig.~\ref{fig1}.
}\label{fig2}
\end{center}
\end{figure}

For each possible state at time $t$, which may give rise to four
possible states at time $t+1$ with certain probability, the
transition probability trees for node $i$ are illustrated in
Fig.~\ref{fig2}. According to the scheme in Fig.~\ref{fig2}, one can
easily write the Markov Chain Approach (MMCA) equations~\cite{gomez2010discrete,wang2003epidemic} for each state, for example, along
the top branch of the four trees, one can read the probability  $p_{i}^{SU}(t+1)$ of SU
at $t+1$ as:
\begin{eqnarray}\label{5}
p_{i}^{SU}(t+1)&=&p_{i}^{IA}(t)\delta\mu+p_{i}^{IU}(t)\gamma_{i}(t)\mu+p_{i}^{SA}(t)\delta
q_{i}^{U}(t)+p_{i}^{SU}(t)\theta_{i}(t) q_{i}^{U}(t).
\end{eqnarray}

Similarly, the other three MMCA equations can be written as:

\begin{eqnarray}\label{6}
p_{i}^{SA}(t+1)&=&p_{i}^{IA}(t)(1-\delta)\mu+p_{i}^{IU}(t)(1-r_{i}(t))\mu\\
\nonumber&&+p_{i}^{SA}(t)(1-\delta)
q_{i}^{A}(t)+p_{i}^{SU}(t)(1-\theta_{i}(t)) q_{i}^{A}(t),\\
p_{i}^{IU}(t+1)&=&(1-\sigma)\bigg\{p_{i}^{IA}(t)\delta(1-\mu)+p_{i}^{IU}(t)r_{i}(t)(1-\mu)\\
\nonumber&&+p_{i}^{SA}(t)\delta
(1-q_{i}^{U}(t))+p_{i}^{SU}(t)\theta_{i}(t)
(1-q_{i}^{U}(t))\bigg\},\label{7}\\
\nonumber
p_{i}^{IA}(t+1)&=&\sigma\bigg\{p_{i}^{IA}(t)\delta(1-\mu)+p_{i}^{IU}(t)r_{i}(t)(1-\mu)+p_{i}^{SA}(t)\delta
(1-q_{i}^{U}(t))\\
&&+p_{i}^{SU}(t)\theta_{i}(t)
(1-q_{i}^{U}(t))\bigg\}+p_{i}^{IA}(t)(1-\delta)(1-\mu)\\
\nonumber&&+p_{i}^{IU}(t)(1-r_{i}(t))(1-\mu)+p_{i}^{SA}(t)(1-\delta)
(1-q_{i}^{A}(t))\\
\nonumber &&+p_{i}^{SU}(t)(1-\theta_{i}(t))
(1-q_{i}^{A}(t)),\label{8}
\end{eqnarray}
where
$p_{i}^{SU}(t)+p_{i}^{SA}(t)+p_{i}^{IU}(t)+p_{i}^{IA}(t)\equiv1$.
When the system goes to the stationary state, we have
$p_{i}^{SU}(t+1)=p_{i}^{SU}(t)=p_{i}^{SU}$ for SU state and
equivalently for the SU, SA and IU states.

Since the epidemic threshold determines whether the
epidemic can outbreak or die out, it is vital to analyze the
effects of the different parameters on the epidemic threshold $\beta_c$.  As stated in Ref.~\cite{granell2013dynamical}, near the threshold, the probability of nodes being infected is
very low, i.e., $p_{i}^{I}=\varepsilon_{i}\ll1$. Consequently,
$q_{i}^{A}\approx1-\beta^{A} \sum_j(a_{ji}\varepsilon_{j})$ and
$q_{i}^{U}\approx1-\beta \sum_j(a_{ji}\varepsilon_{j})$. Further
approximate $q_{i}^{A}\approx1$ and $q_{i}^{U}\approx1$ by assuming
$p_{i}^{I}=\varepsilon_{i}\rightarrow0$, and then substitute both of
them into Eq.~(\ref{5}) and Eq.~(\ref{6}), we obtain
\begin{equation}\label{9}
p_{i}^{SU}=p_{i}^{SA}\delta+p_{i}^{SU}\theta_{i}
\end{equation}
\begin{equation}\label{10}
p_{i}^{SA}=p_{i}^{SA}(1-\delta)+p_{i}^{SU}(1-\theta_{i})
\end{equation}

Combing Eqs.~(7--10), then a simple
formula is obtained:
\begin{equation}\label{11}
\mu \varepsilon_{i}=(p_{i}^{SU} \beta+p_{i}^{SA} \beta^{A})
\sum(a_{ji}\varepsilon_{j}).
\end{equation}

With $\beta^{A}=\gamma \beta$,
$p_{i}^{U}=p_{i}^{SU}+p_{i}^{IU}\approx p_{i}^{SU}$,
$p_{i}^{A}=p_{i}^{SA}+p_{i}^{IA}\approx p_{i}^{SA}$ and
$\varepsilon_{i}=p_{j}^{IU}(t)+p_{j}^{IA}(t)\ll1$, then
Eq.~(\ref{11}) can be rewritten as:

\begin{equation}\label{12}
\sum[(1-(1-\gamma) p_{i}^{A}) a_{ji}-\frac{\mu}{\beta}
\delta_{ji}]\varepsilon_{j}=0,
\end{equation}
here $\delta_{ji}=1$ if $i=j$; otherwise, $\delta_{ij}=0$.

Defining matrix $H$ with elements:
\begin{equation}\label{13}
h_{ji}=(1-(1-\gamma) p_{i}^{A}) a_{ji},
\end{equation}
 Eq.~(\ref{12}) can be read as
\begin{equation}\label{14}
H \bm{\varepsilon}=\frac{\mu}{\beta}\bm{\varepsilon},
\end{equation}
where
$\bm{\varepsilon}=(\varepsilon_1,\varepsilon_2,\cdots,\varepsilon_N)^{T}$
with $T$ be the vector transportation.

The non-trivial solutions of Eq. (\ref{14}) are eigenvectors of $H$,
whose eigenvalues are equal to $\mu/\beta$. Therefore, the onset of
the epidemics is given by the largest eigenvalue of $H$,\emph{i.e.},
$\bigwedge_{max}(H)$,

\begin{equation}\label{15}
\beta_{c}=\frac{\mu}{\bigwedge_{max}(H)}.
\end{equation}

\subsection{Numerical simulations}
To verify our theoretical results, as in Ref.~\cite{granell2013dynamical}, we build a configuration network with degree distribution $P(k)\sim k^{-2.5}$ and network size
$N=2000$ as the contact network, and for the information
network, which is generated by adding 800 extra random links in the
contact network. $a_{ij}$ and $b_{ij}$ represent the adjacency
matrices of the contact network and the information network,
respectively. All simulation results are obtained by averaging 20 realizations.

\begin{figure}
\begin{center}
\includegraphics[height=60mm,width=130mm]{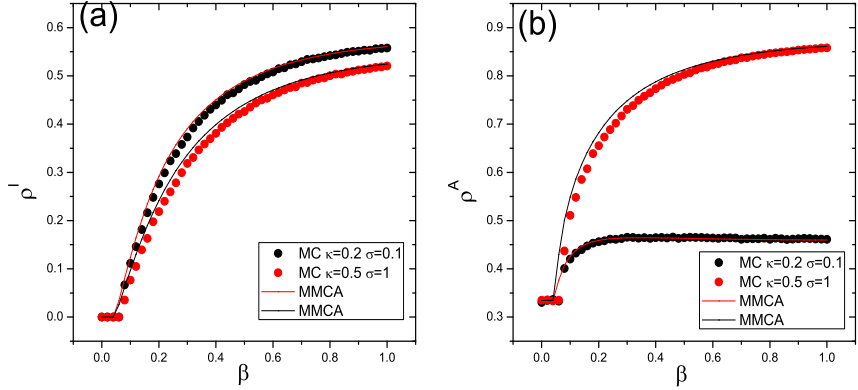}
\caption{ Comparison of MMCA(solid line) with Monte Carlo
simulation(dotted line) for $\lambda=0.15$, $\gamma=0$, $\delta=0.6$
and $\mu=0.4$. The fraction of infected (a) and aware (b) nodes as a
function of the infectivity parameter $\beta$ for two different
conditions of the parameters $\kappa$ and $\sigma$.}\label{fig3}
\end{center}
\end{figure}

We first compare the results from MMCA with Monte-Carlo simulation in
Fig.~\ref{fig3} to check the effectiveness of our analysis based on
MMCA, from Fig.~\ref{fig3}, one can observe that the results based
on the two approaches are in good agreement. So in the next figures, our main
results are obtained from MMCA.

Then we investigate the effects of the two main parameters of the
model---$\kappa$ and $\sigma$
on the epidemic threshold and the density of infected individuals. Here, we will present the
results for $\gamma=0$, meaning that $\beta^{A}=0$ and $q^{A}=0$. Obviously, once the value of $\gamma$ is increased, the epidemic threshold is decreased and the density of infected individuals in enhanced correspondingly.

\begin{figure}
\begin{center}
\includegraphics[height=60mm,width=130mm]{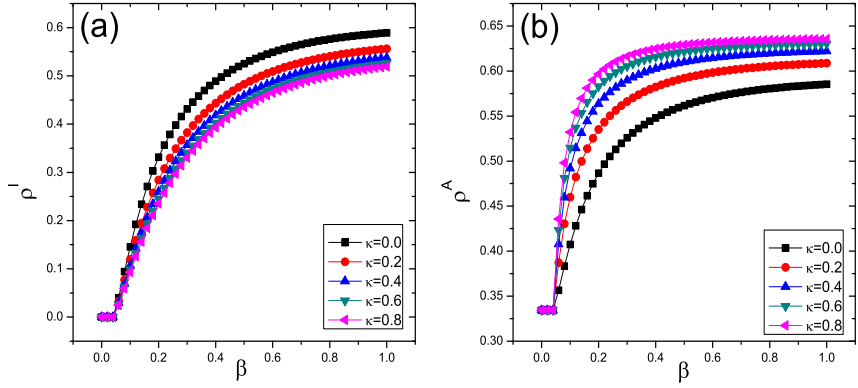}
\caption{Fraction of infected (a) and aware (b) nodes as a function
of the infectivity parameter $\beta$ for different values of the
parameter $\kappa$.
Here $\lambda=0.15$, $\gamma=0$, $\delta=0.6$, $\mu=0.4$ and
$\sigma=0.4$.}\label{fig4}
\end{center}
\end{figure}

Fig.~\ref{fig4} plots the density of infection [$\rho^{I}$, see Fig.~\ref{fig4}(a)] and aware
individuals [$\rho^A$, see Fig.~\ref{fig4}(b)] as a function of $\beta$ for different values of
$\kappa$, respectively.  Observing
the figure, one can see that though the larger value of $\kappa$ can cause more individuals become aware and reduce the density of infection. However, one can find that the increasing of $\lambda$ has no influence on the epidemic threshold. The result is remarkably different from many previous results
which claim that the local information-based behavioral response in the \emph{single-layer network} can alter the epidemic threshold. How to understand this nontrivial result? Since the UAU awareness dynamic is the same to the SIS epidemic process. When the epidemic has not broken out, the density of awareness in information network (\emph{i.e.}, $\rho^A$) is only determined by the transmission rate of awareness, $\lambda$, recovery rate $\delta$ and the structure of information network, but is not related to the value of $\kappa$.  Namely, near or below the epidemic threshold point, increasing the value of $\kappa$ only means that the initial number of the aware individuals is increased, which cannot affect the density of aware individuals at stationary state. In this case, the value of $p_{i}^{A}$ is independent of the value of $\kappa$ or $\sigma$, which gives rise to the same value of $\rho^{A}$ [see Fig.~\ref{fig4}(b)]. Thus, according to Eqs.~(\ref{13}) and (\ref{15}), the epidemic threshold $\beta_c$ is invariable owing to the same value of $\rho^A$.  One should note that,
for the case of single-layer network, the local information-based behavioral response can \emph{directly} reduce the transmission rate of epidemic, leading to the change of the epidemic threshold. For our model, the self-awareness behavior first diffuses through the information network, and then the epidemic process happens in contact network. Thus, the effect of the awareness behavior on the transmission rate is \emph{indirect}.

\begin{figure}
\begin{center}
\includegraphics[height=60mm,width=130mm]{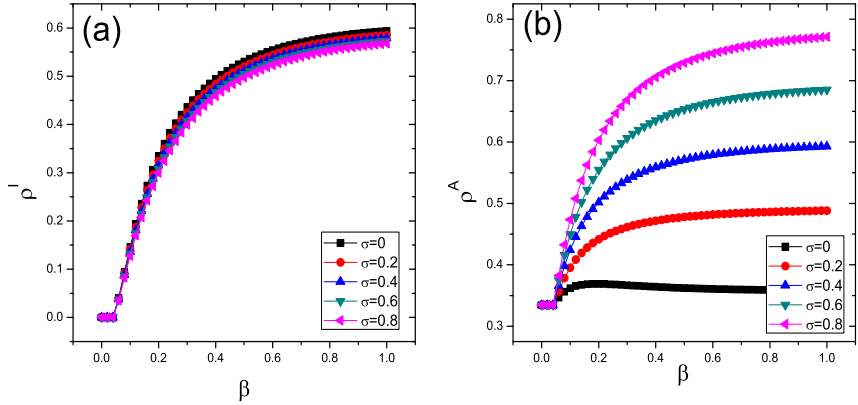}
\caption{Fraction of infected (a) and aware (b) nodes as a function
of the infectivity parameter $\beta$ for different values of the
parameter $\sigma$. Here
$\lambda=0.15$, $\gamma=0$, $\delta=0.6$, $\mu=0.4$ and
$\kappa=0.05$.}\label{fig5}
\end{center}
\end{figure}

The density of infected individuals and aware
nodes as functions of $\beta$ for different values of parameter
$\sigma$ are also shown in Fig.~\ref{fig5}(a) and Fig.~\ref{fig5}(b), respectively. One can see that, similar to Fig.~\ref{4}, varying the value of $\sigma$ has no effect on the epidemic threshold. However, differ to the above case, from Fig.~\ref{fig5}(a) we find that the value of $\sigma$ also has negligible effect on the density of infection, even in the extreme cases where infected unaware individuals remain unaware of its sickness
($\sigma=0$) or certainly become aware of it ($\sigma=1$). The result in Fig.~\ref{fig5} is in accord with the Fig.~3 in Ref.~\cite{granell2014competing}.

\begin{figure}
\begin{center}
\includegraphics[height=130mm,width=130mm]{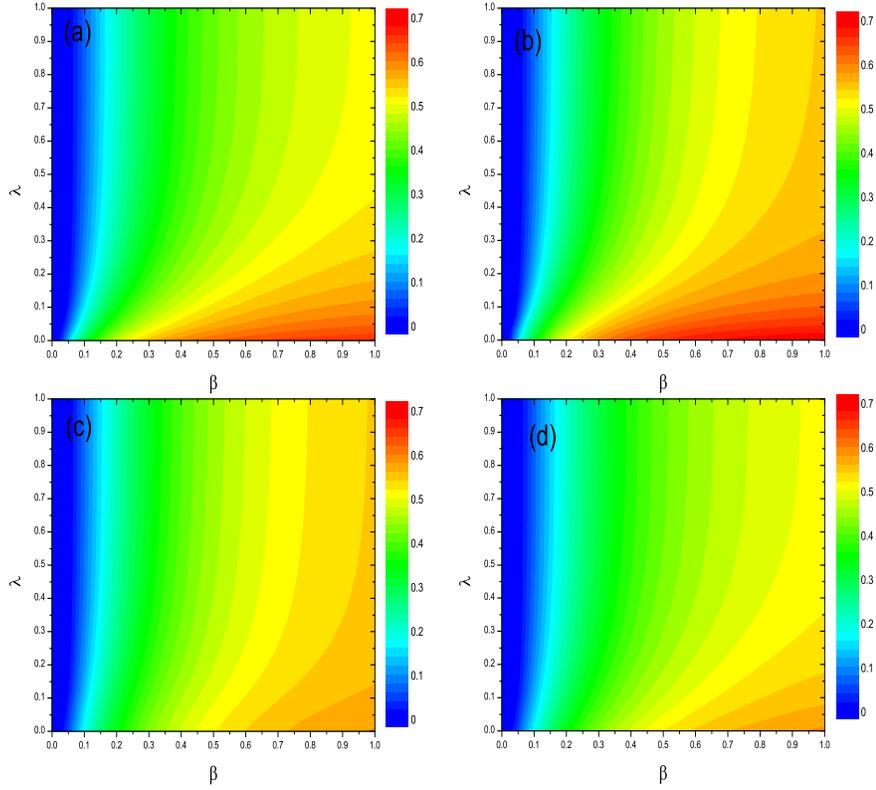}
\caption{The fraction $\rho^{I}$ of infected individuals in the
stationary state. Full phase diagram $\lambda-\beta$ for the same
multiplex described before Where $\gamma=0$, $\lambda=0.15$,
$\delta=0.6$ and $\mu=0.4$. (a): $\kappa=0$, $\sigma=1$; (b):
$\kappa=0$, $\sigma=0.2$; (c): $\kappa=0.2$, $\sigma=0.2$; (d):
$\kappa=0.2$, $\sigma=1$.}\label{fig6}
\end{center}
\end{figure}

In order to systematically study the effects of $\kappa$ and $\sigma$ on the $\rho^I$, we further explore the full phase
diagram ($\lambda-\beta$) in Fig.~\ref{fig6}. Overall, we can see that $\rho^I$ is not influenced by $\lambda$ when $\beta$ is smaller than the epidemic threshold, since epidemic will die out by itself. Once $\beta$ overpasses the epidemic threshold, $\rho^I$ decreases with $\lambda$ for different values of $\kappa$ or $\sigma$. More specifically, by comparing
Fig.~\ref{fig6}(a) with Fig.~\ref{fig6}(b) (or comparing Fig.~\ref{fig6}(c) with
Fig.~\ref{fig6}(d)), we can see that $\rho^I$ is not remarkably influenced by the value of $\sigma$. Likewise, by comparing the
Fig.~\ref{fig6}(a) with Fig.~\ref{fig6}(d) (or comparing
Fig.~\ref{fig6}(b) with Fig.~\ref{fig6}(c)), $\rho^I$ decreases with $\kappa$, especially for the large value of $\beta$.

\section{Conclusions and discussions} \label{sec:discussion}

Recognizing that, on the one hand, outbreak of an epidemic through a physical-contact network can trigger the spreading of information awareness through other different channels, such as on-line social networks, mass media; on the other hand, an individual can not only be informed by other aware neighbors but also can become self-awareness once some friends in contact network are infected.  By introducing the \emph{self-awareness} mechanism for susceptible individuals, we have investigated interplay between the spreading of epidemic and the diffusion of awareness based on the framework of the multiplex networks. We mainly studied the two parameters $\kappa$ and $\sigma$ characterizing the self-awareness probability of susceptible individuals and infected individuals, respectively. Analysis based on the Markov chain approach as well as the extensive computations reveal that the density of infection can be reduced once the two parameters are increased, however, we found that the impact of self-awareness behavior for susceptible individuals on inhibiting the spreading of epidemic is much better than the self-awareness of the infected individuals, since self-awareness from susceptible individuals can directly reduce their probabilities of being infected. Meanwhile, we found that the self-awareness behavior cannot alter the epidemic threshold no matter of the local or global information, which are in stark contrast with the results obtained from the single-layer networks.

The challenges of studying the intricate interplay between social and biological contagions in human populations are generating interesting science. In this work, we consider the effects of the \emph{self-awareness} behavior based on the multiplex networks on the density of infection and the epidemic threshold, our result implies that the conclusions obtained from single-layer networks may need to be re-examined when they are extended to multiplex networks.

\section{Acknowledgments}

This work is funded the National Natural Science Foundation of China
(Grant Nos. 61473001,11331009) and the Doctoral Research
Foundation of Anhui University (Grant No. 02303319).

\section{Apendix: Global information-based awareness}
In recent work, Granell \emph{et~al.} considered the effect of the mass media on the epidemic process and awareness diffusion~\cite{granell2014competing}. In the model, each individual
becomes aware with probability $m$ by assuming that they are informed by a broadcast or mass media. Thus, it can be regarded as a global information-based awareness. One questionable point is that the probability of being awareness $m$ is irrelevant to the density of infection. As a result, even the epidemic is almost eliminated, individuals also have the fix probability of being aware. In reality, becoming awareness often means that individuals need to take some protective measures, such as, washing hands, wearing masks or reducing outgoings. These measures indicate certain inconveniences or some cost~\cite{haifeng2013braess,fu2011imitation}. Thus, a more realistic situation is that the probability of being awareness should be related to the density of infection. To mimic this case, here we assume the probability of aware from global information is given as: $mI(t)$ with $I(t)$ is the density of infection at time $t$, which indicates that the probability of being awareness adaptively varies with the density of infection.

For this case, we only need to slightly change the
local model described in subsection ~\ref{analysis}. We only need to change $\gamma_{i}(t)$ and
$\theta_{i}(t)$ as follows:

\begin{equation}\label{16}
    \gamma_{i}(t)=\prod_{j}(1-b_{ji}p_{j}^{A}(t)\lambda)(1-mp^{I}(t)),
\end{equation}
\begin{equation}\label{17}
    \theta_{i}(t)=\prod_{j}(1-b_{ji}p_{j}^{A}(t)\lambda)(1-a_{ji}p_{j}^{I}(t)\kappa)(1-mp^{I}(t))
\end{equation}

\begin{figure}
\begin{center}
\includegraphics[height=90mm,width=100mm]{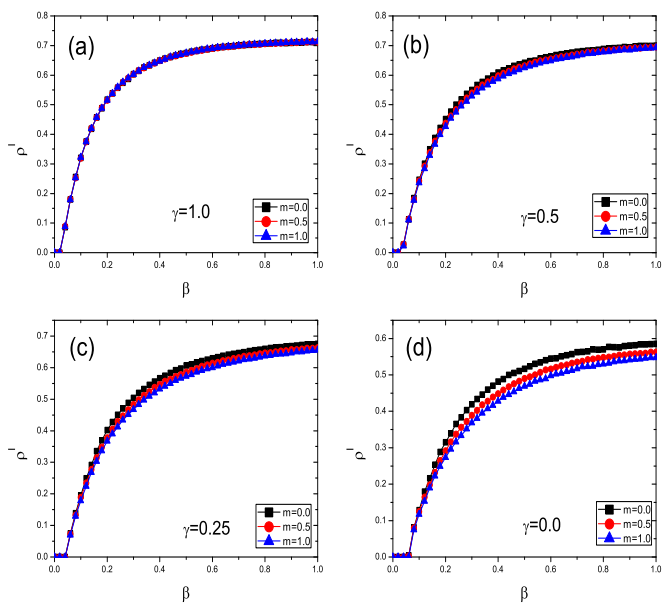}
\caption{The density of infected individuals as a function of $\beta$ for different values of $m$. Here $\lambda=0.3$, $\sigma=0.5$
$\delta=0.6$, $\mu=0.4$ and $\kappa=0.0$. (a): $\gamma=1$; (b):
$\gamma=0.5$; (c): $\gamma=0.25$; (d): $\gamma=0$.}\label{fig7}
\end{center}
\end{figure}

Similarly to the above analysis, we can get that the epidemic threshold is still determined by Eq.~(\ref{15}),\emph{ i.e.}, the epidemic threshold is also independent of the value of $m$, which is different from the result in Ref.~\cite{granell2014competing}. The result is verified by Fig.~\ref{fig7} for different values of $\gamma$.

\end{document}